Letter to the editor

# Hybrid cold and hot-wall chamber for fast synthesis of uniform graphene

*Hadi Arjmandi-Tash[1], Nikita Lebedev[2], Pauline van Deursen[1], Jan Aarts[2], and Grégory F. Schneider[1]\**

*Leiden University, Faculty of Science, Leiden Institute of Chemistry, Einsteinweg 55, 2333 Leiden, The Netherlands*

*Leiden University, Faculty of Science, Leiden Institute of Physics, Niels Bohrweg 2, 2333 Leiden, The Netherlands*

**Abstract**

We introduce a novel modality in the CVD growth of graphene which combines the cold-wall and hot-wall reaction chambers. This hybrid mode preserves the advantages of a cold-wall chamber as the fast growth and low fuel consumption, but boosts the quality of the growth towards conventional CVD with hot-wall chambers. The synthesized graphene is uniform and monolayer. The electronic transport measurements shows great improvements in charge carrier mobility compared to graphene synthesized in a normal cold-wall reaction chamber. Our results promise the development of a fast and cost-efficient growth of high quality graphene, suitable for scalable industrial applications.

**Introduction**

Cold-wall chambers (CWC) are of advantageous for the growth of graphene as they allow fast synthesis. Cold-wall chambers can be constructed compact; the small size of the reaction chamber allows lower gas consumption. The heating energy is selectively used to heat-up the specimen, e.g. copper foil in contact with the hot stage; hence the energy dissipation is low which minimizes the overall growth costs[1].

On the negative side, however, the knowledge and experience about the chemical vapor deposition (CVD) of graphene with CWC is very limited. Compared to HWC, successful reports for the growth with CWC are rare[1–4] which accounts for a general sense of distrust in the community regarding the utilization of the CWCs. This manuscript studies CVD growth of graphene in a CWC; we offer solutions to improve the quality of the synthesized graphene in the CWC which include improving the growth parameters and adopting the growth principal of the HWC, i.e. hybridizing CWC and HWC. The modifications are successful to boost-up the uniformity and the electronic transport properties of the synthesized graphene to be comparable with graphene grown in conventional HWCs.

**Comparison of the CWC and HWC**

Figure 1 compares typical CWC and HWC setups. In the HWC, the heating elements are placed outside the chamber tube; radiation of the thermal energy via the transparent quartz tube, heats-up the specimen (copper foil) placed inside the tube (see the inset figure 1-b). Typically, the heating elements are surrounded by a large block of insulating materials to minimize the energy dissipation to the environment. This block, however, acts as thermal mass which delays both the heating (to start the growth) and cooling of the chamber (at the end of the process). In a CWC, however, the specimen is placed directly on a resistively heated stage inside the chamber (left bottom inset Figure 1-a). In typical designs, the size of the heating stage could be as small as the size of the specimen with no insolating materials required, which makes fast processing possible. The greatest difference between the counterpart chambers is that uniform radiation in a HWC provides a large (compared to the size of the specimen) heating zone with a uniform temperature whereas there is a huge thermal gradient between the hot stage (~1000 °C) and the cold walls (~few tens of °C) during the operation of the CWC (bottom inset Figure 1-a and b).

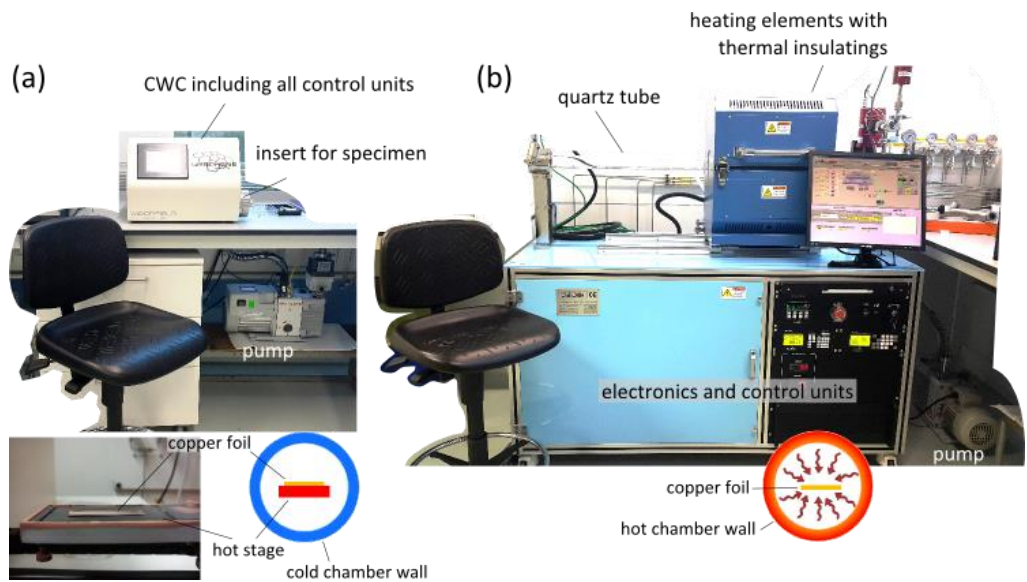

Figure 1: Cold- versus hot-wall reaction chambers for the growth of graphene
a) Photograph of a commercially available cold-wall chamber (nanoCVD-8G, Moorfield Nanotechnology): The main unit is of 40.5 cm × 41.5 cm × 28 cm dimensions and weighs 27 kg. The bottom-left inset shows the hot-stage (4.0 cm × 2.5 cm) hosting a copper foil. The bottom-right inset schematically shows the chamber.
b) Photograph of a commercially available hot-wall chamber (planarGROW-2B, plamarTECH): The unit is of 1.75 m × 1.60 m × 0.75 m and weighs ~200 kg. The schematics shows the radiative heat transfer to the specimen in the chamber.

**Growth of graphene in a CWC**

Figure 2 characterizes graphene grown in a CWC. For this growth, we adopted a recipe similar to what has been developed earlier[1,2,5] and includes: i) heating the copper foil to 1035 °C, ii) annealing for 10 minutes and iii) growth for 10 minutes ($CH_4$ to $H_2$ ratio of 7:2, gas purity grade 6.0, chamber pressure of few mbar). The synthesized graphene covers the surface of the copper thoroughly, yet it suffers from several imperfections (summarized in Table 1):

The presence of multilayer areas is the first imperfection, evident as rounded or linear patches of different contrasts in optical micrograph in Figure 2-a: Indeed those multilayer islands grow due to the high population of the defect sites on the copper foil and with the presence of the excessive carbon precursors[6]. An elongated annealing (up to one hour) lowers the defect sites by improving the surface quality of the copper. Lowering the $CH_4/H_2$ ratio minimizes the excessive carbon precursors. We note that much lower $CH_4/H_2$ ratio of 2 sccm/1000 sccm has achieved a uniform monolayer coverage[6] in a HWC process.

Local crystalline defects manifested by the huge D peak in the Raman spectra is the second imperfection (figure 2-b). Impurities in the utilized gases and/or the oxidation during transfer to $SiO_2$ wafer are among potential sources of the D peak in CVD graphene. Improving the quality of the resources (using higher purity gases) and optimizing the transfer process helps to minimize the flaw.

Heterogeneous growth is another imperfection which is evident by dissimilar Raman spectra recorded at different spots of the sample. The imperfection persisted even after the elongated annealing to improve the uniformity of the copper foil. The long quartz tubes used in the HWC ensures laminar and fully developed flow of gases before reaching the copper[7]; the absence of such a "guide" in short reaction chamber may cause local eddies and non-uniform stream. The huge thermal gradient between the hot stage and the walls of the chamber and non-uniform heating due to the small heating zone could be other sources of the inhomogeneity. Indeed this imperfection can be viewed as a strong and intrinsic side effect of the compact and energy-efficient design of the CWCs. A solution to lower the drawbacks of the design while preserving the benefits includes covering the stage with a quartz plate, while leaving a gap of ~2 mm for the flow of gases (Figure 2-c). The benefit is twofold: The flow of the gases through the gap is inside the boundary layer associated with the cap hence is uniform. Additionally, within this design, the heat radiated out from the hot stage during the growth is reflected back to the copper foil by the shiny surface of the quartz; hence inside the gap, a small reaction chamber which is a hybrid of CWC and HWC (C/HWC) with a uniform temperature develops.

Figure 2-d and e show the optical micrograph of a selected area and multiple Raman spectra recorded at different spots of the graphene synthesized with the improved recipe in the C/HWC. Obviously, the modifications improved the uniformity of the growth and eliminated the multilayer patches. Although there is still a D peak detectible in Raman spectra, the lowered $I_D/I_G$ ratio indicates the improved crystalline structure. The inset figure 2-e focuses on a selected spectra between 1200 cm$^{-1}$ and 1700 cm$^{-1}$. D, G and D' peaks are clearly visible and deconvoluted by means of Gaussian fits. We estimated $I_D/I_{D'}$ = 2.75, close to the value reported for the grain boundary defects[8] indicating that the synthesized graphene suffers from a high population of the grain boundaries, i.e. small grains. The complementary electron diffraction pattern of a suspended graphene sample (figure 2-f) shows the presence of regions without preferred lattice orientation, i.e polycrystalline graphene. Note that the Raman spectra with similar $I_D/I_{D'}$ can be identified in early reports with CWC[1]; Indeed small grains is the characteristics of graphene growth in the CWC.

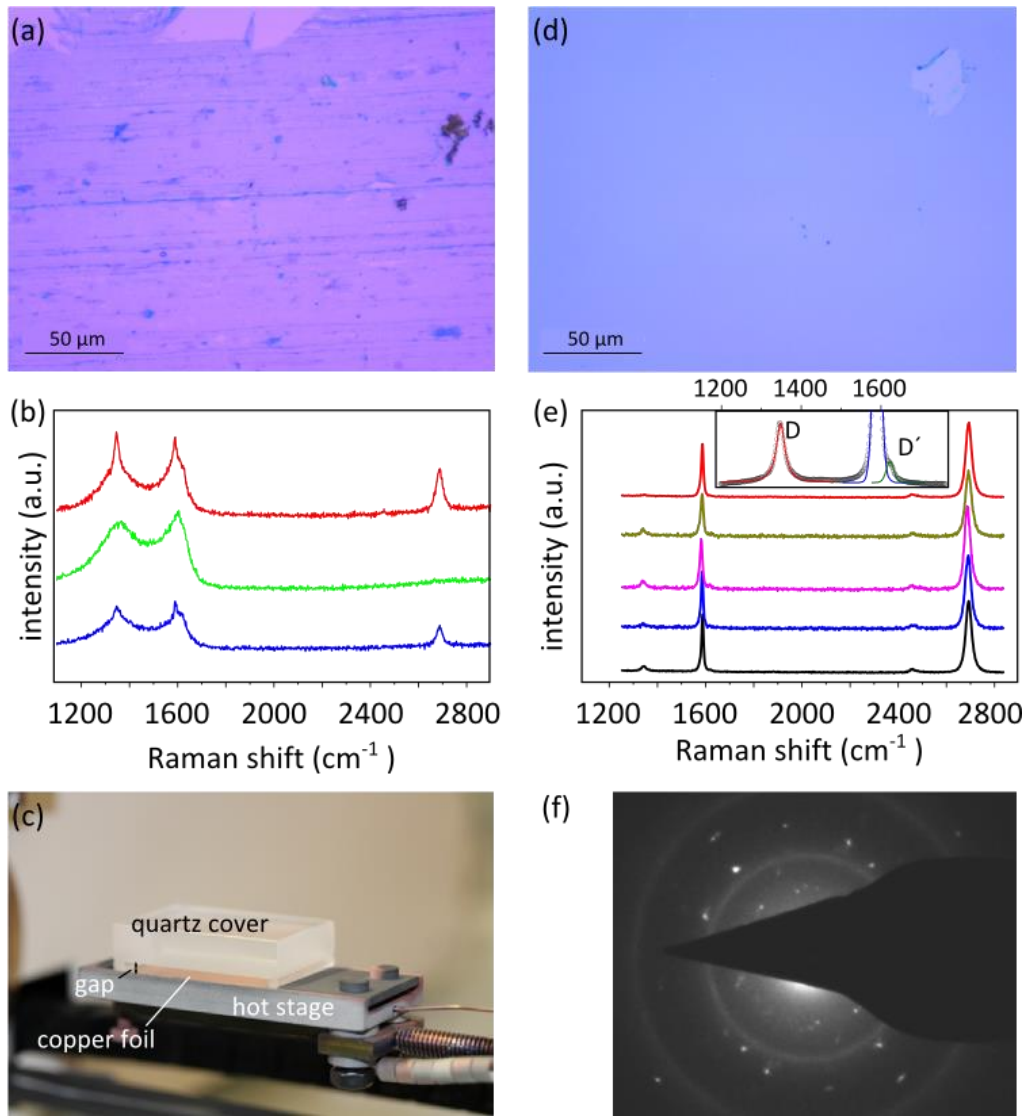

Figure 2: Characterization of graphene synthesize in a CWC
a) Optical micrograph illustration of a graphene sheet synthesized via a conventional recipe for the growth in a CWC (detailed in the text), transferred onto a SiOx/Si wafer
b) Typical Raman spectra corresponding to different spots on graphene in a.
c) Photograph illustrating the technique to turn a CWC into hybrid C/HWC
d) Optical micrograph illustration of a graphene sheet synthesized via an improved recipe (detailed in the text), in a hybrid C/HWC, transferred onto a SiOx/Si wafer
e) Typical Raman spectra corresponding to different spots on graphene in d, the inset details a frequency window close to D , G and D' peaks.
f) Typical diffraction pattern corresponding to free standing graphene grown in the hybrid C/HWC, recorded by diffraction mode transmission electron microscopy

Table 1: Imperfections in chemically synthesized graphene in a CWC

| imperfection | origin | solution |
| --- | --- | --- |
| multilayer areas | presence of the defect sites on Cu, | increasing the annealing duration |
|  | excessive carbon precursor | lowering $CH_4/H_2$, shortening the growth |
| Raman D peak | contaminations in the supplies | using higher quality supplies |
|  | oxidation during transferring | improving the transfer |
| heterogeneous growth | non-uniform heating | hybridizing the CWC and HWC |

**Electrical measurement results**

We characterized the electrical performance of the graphene samples grown via our hybrid C/HWC. Black data points in Figure 3-a illustrate the gate-dependent resistivity of a selected sample measured at room temperature. The continuous line shows the result of the best fitting with the existing model for the conductivity (σ) of graphene[9]: $\sigma^{-1} = (ne\mu_c + \sigma_0)^{-1} + \rho_s$. Here $\mu_c$ is the density-independent charge carrier mobility, e is elementary charge, $\sigma_0$ is the residual conductivity at the Dirac point and $\rho_s$ is the contribution of short-range crystalline defects on the total resistivity. Additionally, n is the charge carrier density, estimated considering the parallel-plate capacitance model across the oxidized silicon layer ($\varepsilon_r = 3.9$). Figure 3-b compares the extracted room temperature $\mu_c$ of several graphene samples grown via hybrid C/HWC and conventional CWC with this fitting. The samples grown using our hybrid C/HWC exhibited an average mobility of $1.5\times10^3$ $cm^2/V.s$, showing ~27% improvement with respect to the samples grown via the conventional recipe ($1.2\times10^3 cm^2/V.s$). The improvement is attributed to the uniform crystalline structure of the hybrid C/HWC devices.

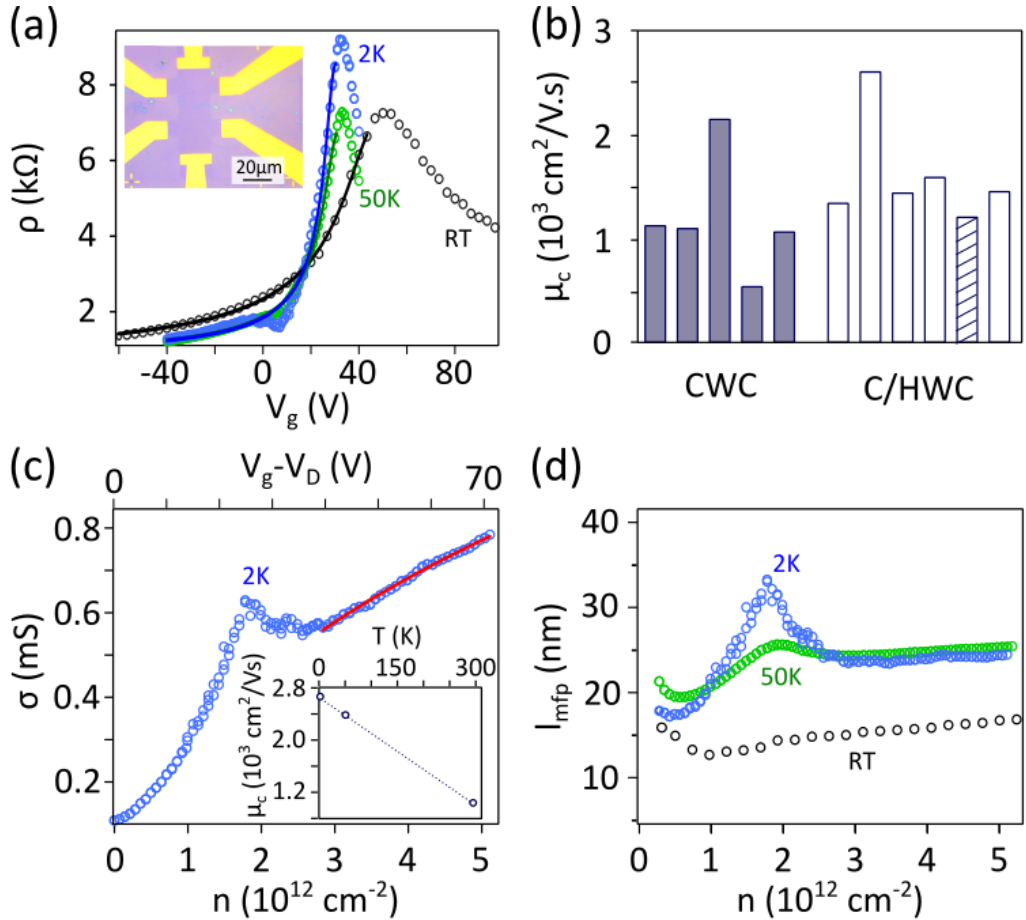

Figure 3: Electric transport properties of graphene grown via hybrid C/HWC
a) Gate dependent electrical resistivity of a sample measured at different temperatures: The continuous lines are the best fittings with the model of the conductivity of graphene, discussed in the text. An optical micrograph of the sample is presented in the inset figure.
b) Mobility of different graphene samples synthesized via conventional CWC and hybrid C/HWC; the hatched sample is the one presented in (a).
c) Conductivity of the same sample in a, measured at 2K: continuous line is the best fitting with the mid-gap states model. $V_D$ refers to the gate voltage at the Dirac point. The inset plots the mobility of the sample at different temperatures. The dotted line is guide to the eye.
d) Density dependent mean free path of the charge carriers of the sample in a, at different temperatures

Cooling down the sample lowers the phonon scattering; hence $\mu_c$ improves and reaches 2700 $cm^2/V.s$ at 2K (inset Figure 3-c); the measured mobility is slightly lower than typically reported values for CVD graphene in HWC[10] which may highlight the effect of the grain boundaries in scattering the charge carriers.

Analysis of the field dependent conductivity of the sample at low temperatures reveals the characteristics of the defects ($R_0$ the size and $n_d$ the density) in the graphene lattice via the "Mid-gap states" model[11]: $\sigma = \frac{2e^2}{h} \frac{k_F^2}{\pi n_d} [\ln(k_F R_0)]^2$. Here, $k_F = \sqrt{\pi n}$ is the Fermi wave vector of graphene. The continuous line in Figure3-c is the best fitting of the conductivity with this model.

Due to the short effective range of the crystalline defects, their scattering effect is considerable only with the high population of the charge carriers (i.e. far from the Dirac point)[9] hence close to the Dirac point the model ceases to follow the experimental results. Table 2 summarizes the characteristics of the defects achieved by this fitting. For the sake of comparison, we included the results reported earlier for CVD graphene grown via conventional HWC and the estimation for exfoliated graphene[12].

Table 2: Characterization of the crystalline defects in different graphene samples

| sample | $n_d$ [cm$^{-2}$] | $R_0$ [Å] |
|---|---|---|
| hybrid C/HWC-CVD | $2.1 \times 10^{12}$ | 3.0 |
| HWC-CVD | $2.7 \times 10^{12}$ | 1.3 |
| exfoliated graphene | $\leq 1 \times 10^{11}$ | 1.4 |

Crystalline defect of different types including vacancies, cracks, or grain boundaries contribute in the estimated $R_0$ and $n_d$. Particularly the high population of grain boundaries (with typical sizes larger than single vacancies) in hybrid C/HWC graphene raised the *average* size of the defects beyond HWC-CVD graphene. The density of the defects of both hybrid C/HWC and conventional HWC is approximately one order of magnitude higher than that for an exfoliated graphene which justifies the poorer transport properties of CVD graphene.

Mean free path of the charge carriers are estimated using $l_{mfp} = (h/2e)\mu_{FE}\sqrt{n/\pi}$ where $\mu_{FE} = \sigma/en$ is the field effect mobility of the charge carriers. Figure 3-d plots the carrier density dependent $l_{mfp}$ at different temperatures. By cooling the sample below the room temperature (down to 50 K), the reduction of the phonon scattering elongates the mean free path. Further reduction of the temperature, however, does not affect the $l_{mfp}$, particularly at higher carrier density, $l_{mfp}$ saturates about 25nm which can be attributed to the trapping of the carriers inside graphene grains.

**Conclusion**

We presented a systematic study of the CVD growth of graphene in a cold wall chamber. We identified the important imperfections of the grown of graphene and proposed solutions to eliminate them. Particularly a simple technique can turn the CWC into a hybrid C/HWC which considerably improves the uniformity of the growth and charge carrier mobilities. Small grain size remains an important characteristics and a challenge for the graphene synthesized in a CWC and hybrid C/HWC which limits the transport properties of the graphene.

**Acknowledgements**

The work leading to this article has gratefully received funding from the European Research Council under the European Union's Seventh Framework Programme (FP/2007-2013)/ERC Grant Agreement n. 335879 project acronym 'Biographene' and the FP7 funded DECATHLON Grant agreement n. 613908 'DEvelopment of Cost efficient Advanced DNA-based methods for specific Traceability issues and High Level On-site applicatioNs', and the Netherlands Organization for Scientific Research (Vidi 723.013.007).